\documentclass[preprint,12pt]{elsarticle}

\usepackage{amssymb}
\usepackage{amsmath}
\begin{document}

\begin{frontmatter}



\title{Chiral Doubling, Renormalization Group Fixed Points, and Dense Matter Equations of State}


\author{Chihiro Sasaki} %% Author name

\affiliation{organization={Institute of Theoretical Physics, University of Wroclaw},%Department and Organization
            addressline={plac Maksa Borna 9},
            city={Wroclaw},
            postcode={PL-50204},
            country={Poland}}
\affiliation{organization={International Institute for Sustainability with Knotted Chiral Meta Matter (WPI-SKCM$^2$), Hiroshima University},%Department and Organization
            addressline={1-3-1 Kagamiyama},
            city={Higashi-Hiroshima},
            postcode={739-8531},
            country={Japan}}

%% Abstract
\begin{abstract}
We review a top-down fixed-point-extrapolation paradigm as a unified effective field theory framework for hadronic spectroscopy,
dense baryonic matter, and compact star physics.
By formulating our effective Lagrangian directly at renormalization group (RG) fixed points and introducing minimal symmetry breaking,
we extrapolate the theory back to physical environments.
In the vacuum, this framework naturally reproduces the heavy-light meson parity-doubling spectrum via light vector meson loops.
In dense matter, the interplay between the chiral-invariant nucleon mass $m_0$ and walking vector couplings reconciles gravitational-wave constraints
with $2M_\odot$ massive neutron stars.
Generalization to a quark-hadron hybrid approach further illuminates sequential deconfinement, core stability, and baryon number fluctuations, demonstrating how underlying RG fixed points dictate hadronic dynamics from vacuum to neutron star interiors.
\end{abstract}

\end{frontmatter}


%%%%%%%%%%%%%%%%%%%%%%%%%%%%%%%%
\section{Introduction}
%%%%%%%%%%%%%%%%%%%%%%%%%%%%%%%%

The construction of effective field theories (EFTs) anchored to the fundamental symmetries of Quantum Chromodynamics (QCD) provides a powerful non-perturbative framework for describing hadronic phenomena across diverse energy and density regimes~\cite{Brown:2001nh,Holt:2014hma,Rho:2024ihu,Rho:2025rqf,Rho:2025mns,Fukushima:2013rx}.
A particularly insightful strategy within this paradigm is to define the theory at a renormalization group (RG) fixed point,
where the system exhibits a highly symmetric or simplified phase, and subsequently extrapolate the dynamics to the physical world
by introducing minimal and systematic departures from the fixed point.
By constraining the Lagrangian with the algebraic structure at the fixed point,
one can drastically reduce theoretical ambiguities and construct an effective description that remains predictive even in the broken phase.

In this tribute, we review this top-down fixed-point-extrapolation paradigm as a unified approach to hadronic spectroscopy and dense matter physics.
Rather than relying on conventional bottom-up models that start from the spontaneously broken vacuum and track the system toward chiral restoration
under extreme conditions,
we formulate our effective Lagrangian directly at the relevant RG fixed points,
such as the Vector Manifestation (VM)~\cite{Harada:2000kb,Harada:2003jx,Harada:2001it,Harada:2001qz}.
By introducing the minimal breaking parameters, we extrapolate the theory back to physical environments.

This top-down strategy proves remarkably successful across a broad spectrum of hadronic phenomena.
Applied first to vacuum spectroscopy, expanding the effective action around the VM fixed point naturally generates the mass splitting $\Delta M$
between negative-parity $H(0^-, 1^-)$ and positive-parity $G(0^+, 1^+)$ heavy-light doublets as a direct consequence of explicit departures
from the fixed point~\cite{Harada:2003kt}.
Scaled linearly by $\langle \bar{q}q \rangle$ and combined with crucial quantum loop enhancements from light vector mesons, this framework quantitatively reproduces the observed spectrum at BaBar, CLEO, and Belle.
Extending the same strategy to the baryonic sector, the fixed-point structure readily accommodates the nucleon parity-doublet scheme and the QCD trace anomaly within dilaton-implemented Hidden Local Symmetry (dHLS)~\cite{Sasaki:2011ff,Paeng:2011hy}.
This provides a systematic mechanism to disentangle the chiral-invariant mass $m_0$ of nucleons from condensate-driven mass generation, while in-medium $U(2)$ breaking naturally preserves the short-range vector repulsion essential for dense nuclear matter~\cite{Paeng:2013xya}.

Furthermore, extrapolating these constrained couplings to extreme densities establishes a direct link to the equation of state (EoS) of neutron star matter. This unified framework seamlessly reconciles the early-stage EoS softening required by gravitational wave constraints with the high-density stiffening necessary to support $2M_\odot$ massive compact stars~\cite{Marczenko:2021uaj,Marczenko:2022hyt}.
The paradigm further illuminates sequential deconfinement transitions and baryon number fluctuations in dense environments~\cite{Marczenko:2018jui,Marczenko:2020jma,Marczenko:2023ohi,Koch:2023oez,Marczenko:2024nge}.
By connecting high-density phase structures directly with low-energy vacuum phenomenology,
this top-down extrapolation offers a comprehensive perspective on how underlying RG fixed points dictate hadronic dynamics
from vacuum spectroscopy to the cores of neutron stars.

%%%%%%%%%%%%%%%%%%%%%%%%%%%%%%%%%%%%%%%%%%%%%%%%%%%%
\section{Chiral Doubling of Heavy-Light Mesons}
%%%%%%%%%%%%%%%%%%%%%%%%%%%%%%%%%%%%%%%%%%%%%%%%%%%%

In this section, as the first concrete implementation of our general strategy, we revisit the chiral doubling of heavy-light mesons~\cite{Nowak:1992um,Bardeen:1993ae,Nowak:2003ra}
in the vacuum based on Ref.~\cite{Harada:2003kt}.
The main objective here goes beyond merely fitting experimental spectrum. Instead, it is to demonstrate how the mass splitting between parity partners acts as a direct and sensitive probe
of the light-quark condensate, the order parameter of spontaneous chiral symmetry breaking, and to reveal the surprisingly dominant and non-trivial role played by light vector mesons
near the vector manifestation (VM) fixed point.

Instead of starting from a Lagrangian defined in the broken phase and driving the system toward chiral restoration through external parameters,
our strategy takes the precise reverse direction: we begin with an assumed structure of chiral symmetry right at its restoration point and ask what it predicts for the broken phase.
To describe heavy-light meson systems, such as $D$-mesons consisting of a heavy quark $Q$ and a light anti-quark $\bar{q}$,
we construct an effective theory incorporating both Heavy Quark Symmetry and Hidden Local Symmetry (HLS).
At the VM fixed point characterized by $(g, a) = (0, 1)$ where the HLS gauge coupling $g$ vanishes and the physical pion decay constant $f_\pi \to 0$ in the chiral limit,
the theory naturally generates a chiral-invariant mass $m_0$ for heavy-light mesons, originating from their interaction with the background light-quark cloud.
As the system moves into the broken phase, spontaneous chiral symmetry breaking manifests itself by inducing an explicit mass splitting term.
Defining the parity-odd doublet $H(0^-, 1^-)$ and parity-even doublet $G(0^+, 1^+)$, the fundamental mass hierarchy is established as
\begin{equation}
M_G - M_H = \Delta M_{\text{bare}}\,,
\end{equation}
where $\Delta M_{\text{bare}}$ represents the bare mass splitting between the two chiral partners.

To pin down this bare parameter from first principles, we perform a Wilsonian matching between the EFT correlators and the Operator Product Expansion (OPE) of QCD at the matching scale $\Lambda_M \sim 1.1\text{ GeV}$.
In the high-energy region, the difference between the scalar and pseudoscalar correlators on the OPE side is dictated almost entirely by the light-quark condensate $\langle \bar{q}q \rangle$.
Equating this directly to the corresponding hadronic correlators yields a remarkably simple matching condition:
\begin{equation}
\Delta M_{\text{bare}} = -\frac{2}{3} \frac{\langle \bar{q}q \rangle}{F_D^2}\,,
\end{equation}
where $F_D$ is the $D$-meson decay constant.
This result establishes an unequivocal link: the bare mass splitting of chiral doublers is directly proportional to $\langle \bar{q}q \rangle$.
It immediately follows that if chiral symmetry is restored at high temperature or high density, the splitting between parity partners must vanish.

When we compute quantum loop corrections to decimated scales where physical masses are measured, the HLS/VM framework reveals a striking dynamics.
Contrary to the conventional belief that pions are the major relevant degrees of freedom near the phase transition,
one-loop calculations including vector mesons show that all power and logarithmic divergences coming from pion loops cancel out completely.
This exact cancellation is a direct consequence of the pairs of states, $\pi$-$\rho_{\text{longitudinal}}$ and $H$-$G$, being exact chiral partners.
Instead, the running of the mass splitting is driven exclusively by (transverse) vector meson loops.
Solving the resulting renormalization group (RG) equation yields a substantial quantum enhancement factor of $C_{\text{quantum}} \approx 1.6$.
With standard input values, the bare splitting $\Delta M_{\text{bare}} \approx 190\text{ MeV}$ is boosted by vector meson loops up to:
\begin{equation}
\Delta M \approx 310 \pm 120 \text{ MeV}\,.
\end{equation}
This quantitatively matches the constituent quark mass scale and reproduces the observed mass differences in open-charm systems,
such as $D_{s0}^*(2317)-D_s(1969)$~\cite{BaBar:2003oey,CLEO:2003ggt,Belle:2003guh}.

This vacuum exercise delivers two crucial messages that define our strategy:
\begin{itemize}
\item
The mass splitting is intrinsically anchored to the quark condensate $\langle \bar{q}q \rangle$, validating chiral doubling as a sharp litmus test for chiral symmetry restoration in hot/dense matter.
\item
Pions alone cannot explain the physical spectrum. The quantum enhancement provided by light vector mesons is indispensable.
This concept becomes even more decisive as we move from vacuum systems to dense baryonic matter and compact star physics.
\end{itemize}

%%%%%%%%%%%%%%%%%%%%%%%%%%%%%%%%%%%%%%%%%%%%%%%%%%%%%%%%
\section{Nucleon Parity Doubling and RG Fixed Points}
%%%%%%%%%%%%%%%%%%%%%%%%%%%%%%%%%%%%%%%%%%%%%%%%%%%%%%%%

Having established the essential role of chiral partners and light vector mesons in the vacuum heavy-light sector, we now extend this framework to dense baryonic matter.
While naive models assume that the nucleon mass drops to zero alongside the melting of the light-quark condensate $\langle\overline{q}q\rangle$,
incorporating the mirror assignment of parity doubling reveals a fundamental structure~\cite{Detar:1988kn}:
a chirally invariant mass $m_0$ that remains intact even when chiral symmetry is restored.

Building upon the series of studies in Refs.~\cite{Sasaki:2011ff},
this parity doublet structure is systematically integrated into dilaton-implemented Hidden Local Symmetry (dHLS).
At extreme densities, dHLS flows toward the Dilaton-Limit Fixed Point (DLFP).
However, as shown in Ref.~\cite{Paeng:2011hy}, assuming exact in-medium $U(2)$ symmetry ($g_{\rho NN} = g_{\omega NN}$) forces both couplings to zero near the fixed point.
Because the iso-scalar $\omega$ meson provides the universal short-range repulsion between nucleons, $g_{\omega NN} \to 0$ would eliminate this repulsive core.
The resulting Equation of State (EoS) would become unphysically soft, causing matter to collapse and directly contradicting the observed existence of stable $2M_\odot$ neutron stars.

Ref.~\cite{Paeng:2013xya} resolved this dilemma by demonstrating that medium effects strongly break $U(2)$ symmetry at high density.
A Wilsonian RG analysis reveals distinct scaling behaviors near the fixed point:
\begin{itemize}
\item
$\rho NN$ coupling runs:
Driven by the VM, $g_{\rho NN}$ drops rapidly toward zero, suppressing the $\rho$-exchange tensor force and leaving the pion tensor force dominant.
\item
$\omega NN$ coupling walks:
Due to $U(2)$ breaking, $g_{\omega NN}$ does not scale at one-loop order. It evolves extremely slowly, staying nearly constant at high density.
\end{itemize}
In-medium $U(2)$ breaking decouples $\omega$ from $\rho$, allowing $g_{\rho NN} \to 0$ while keeping $g_{\omega NN}$ active to ensure dense matter stability.
This persistent $\omega NN$ coupling maintains the short-range repulsion needed to support massive compact stars.
After an initial linear drop, the in-medium nucleon mass $m_N^*$ stops decreasing near the density $n \approx 2n_0$ (the Skyrmion to half-Skyrmion transition) and forms a stable plateau:
\begin{equation}
m_N^* \approx m_0 \sim (0.7\text{--}0.8) m_N\,.
\end{equation}

%%%%%%%%%%%%%%%%%%%%%%%%%%%%%%%%%%%%%%%%%%%%%%%%%%%%%%%%%%%%%%%%%%%%%%%%%%%%%
\section{Dense Baryonic Matter and the Equation of State of Neutron Stars}
%%%%%%%%%%%%%%%%%%%%%%%%%%%%%%%%%%%%%%%%%%%%%%%%%%%%%%%%%%%%%%%%%%%%%%%%%%%%%

Having established the fundamental role of chiral parity doubling and the walking vector-meson couplings in dense baryonic matter,
we now extend this framework to the Equation of State (EoS) of neutron stars (NSs).
A central challenge in modern nuclear astrophysics is the well-known "soft-against-stiff" dilemma:
the EoS must be sufficiently soft at intermediate densities ($\sim 1.5 - 3 n_0$) to satisfy the low tidal deformability extracted from
gravitational wave events such as GW170817~\cite{LIGOScientific:2018cki,LIGOScientific:2018hze},
yet stiff enough at higher densities to support massive compact stars, including the $\approx 2 M_\odot$ pulsars
and NICER radius constraints~\cite{Miller:2019cac,Miller:2021qha}.
Based on the symmetry-driven philosophy described in the previous sections,
our recent studies demonstrate that this non-trivial EoS behavior naturally emerges through the interplay of chiral symmetry restoration
and resonance excitations.

In a pure hadronic framework, the onset of the $\Delta(1232)$ isobar resonance typically induces an unphysically drastic softening of the EoS,
collapsing the maximum stellar mass below the observed limits.
However, when it is introduced as parity doubling, where positive- and negative-parity partners become degenerate while retaining a finite chiral-invariant mass $m_0$, the system exhibits a self-regulating mechanism~\cite{Marczenko:2021uaj,Marczenko:2022hyt}:
As density increases toward partial chiral symmetry restoration, the appearance of $\Delta$ matter drives the required early-stage softening
of the EoS to accommodate the GW170817 tidal deformability constraints.
Subsequently, the persistent chiral-invariant mass $m_0$ together with the walking vector repulsion stiffens the EoS back at higher densities,
naturally pushing the maximum mass beyond $2 M_\odot$.
This crucial finding indicates that a pure hadronic EoS with parity doubling and $\Delta$ resonances can fully satisfy all existing multi-messenger constraints, cautioning that concluding the unequivocal presence of deconfined quark matter in compact star cores may still be premature.

To explore the boundary where quark degrees of freedom become manifest,
we formulated a unified hybrid Quark-Meson-Nucleon (QMN) model~\cite{Marczenko:2018jui,Marczenko:2020jma}.
Departing from conventional two-phase approaches joined by arbitrary Maxwell or Gibbs constructions,
the QMN model incorporates nucleon parity doublets, mesons, and quarks within a single Lagrangian.
Crucial to this formulation is the introduction of an auxiliary background field $b$ (the bag field) that provides statistical confinement.
By modifying the Fermi-Dirac distributions, $b$ acts as an infrared (IR) momentum cut-off for quarks and an ultraviolet (UV) cut-off for nucleons.
This effectively suppresses unphysical thermal/density quark fluctuations at low densities while smoothly suppressing hadronic degrees of freedom
as density increases.
Under cold and charge-neutral NS matter in beta equilibrium, the QMN framework reveals that deconfinement does not occur as a single abrupt boundary. Instead, due to differing electric charges and flavor-dependent couplings, the system undergoes a series of sequential phase transitions:
\begin{itemize}
\item
Chiral restoration in the hadronic sector:
At densities corresponding to stars above $\sim 1.8 M_\odot$, chiral symmetry is restored within the confined nucleon sector.
\item
Onset of $d$ quarks:
Down quarks appear first, giving rise to a mixed phase composed of chirally restored hadrons and $d$ quarks.
\item
Onset of $u$ quarks:
At higher densities, $u$ quarks emerge, ultimately leading to a fully deconfined quark-matter phase.
\end{itemize}
This hierarchy structures the interior of massive neutron stars into four distinct density regimes:
a confined chirally-broken phase, a confined chirally-restored phase, a partially deconfined phase, and a fully deconfined phase.
Notably, the intermediate regime closely mirrors the characteristics of the conjectured quarkyonic phase,
where high-momentum quarks co-exist with a hadronic Fermi surface~\cite{McLerran:2018hbz}.

A key structural outcome of the QMN model concerns the stability of pure quark cores in compact stars.
High-mass stars exceeding $1.8 M_\odot$ are predicted to have the cores made of chirally restored yet pure hadronic matter.
When repulsive vector interactions among quarks are incorporated, as required to maintain EoS stiffness,
the onset of fully deconfined quark matter is pushed to ultra-high densities.
Consequently, the gravitational maximum mass is reached along the hadronic or partially deconfined branch of the mass-radius sequence.
Stars attempting to form a pure deconfined quark core cross the threshold of radial instability and become gravitationally unstable against collapse.
This reveals that stable hybrid stars containing fully deconfined quark cores are unlikely to exist within this theoretical paradigm
unless additional attractive mechanisms, such as diquark condensation (color superconductivity), come into play.

Through these developments, the foundational concepts of chiral doubling, dilaton limit fixed points, and symmetry restoration
continue to provide a robust and predictive guidance for unraveling the mysteries of dense matter in the multi-messenger era.
Furthermore, in the analysis of baryon number fluctuations, a key observable in heavy-ion collision experiments in the high-density and low-temperature regime, we investigated the relationship between nucleon fluctuations per parity and total baryon number fluctuations.
Our recent study utilizing a parity doublet model demonstrates that near the chiral phase transition and the liquid-gas phase transition boundaries of nuclear matter, the fluctuations of positive-parity nucleons (protons), which are directly measured in experiments, do not necessarily reflect the total baryon number fluctuations~\cite{Marczenko:2023ohi,Koch:2023oez,Marczenko:2024nge}
This discrepancy stems from complex correlations between parity partners and the intrinsic dynamics of the parity channels.
Consequently, this provides a vital and indispensable insight for accurately interpreting future experimental data from facilities such as LHC, RHIC-BES and FAIR.

%%%%%%%%%%%%%%%%%%%%%%%%%%%%%%%%%%%%%%%%%%%%%%%%%%%%%%%%%%%%%%%%%%%%%%
\section{Concluding Remarks: A Personal Tribute to Mannque Rho}
%%%%%%%%%%%%%%%%%%%%%%%%%%%%%%%%%%%%%%%%%%%%%%%%%%%%%%%%%%%%%%%%%%%%%%

The core philosophy underlying the work reviewed in this tribute is deeply rooted in the physical intuition and research style of Mannque Rho.
His approach was never merely about fitting data, but about discerning the underlying mathematical and physical structure that governs complex hadronic phenomena.
Over the past twenty-five years, Mannque was far more than a brilliant collaborator.
He was an inspiring mentor and a father-like figure to me.
Maintaining a quiet yet enduring closeness with him allowed me the privilege of observing the very moment that wonderful physical ideas were conceived and brought to life.
He was a genuinely extraordinary theorist and a deeply passionate explorer of physics.

%%%%%%%%%%%%%%%%%%%%%%%%%%%%%%%%%%%
\section*{Acknowledgments}
%%%%%%%%%%%%%%%%%%%%%%%%%%%%%%%%%%%

This work is supported in part by the Polish National Science Centre (NCN) under OPUS Grant No. 2022/45/B/ST2/01527
and the World Premier International Research Center Initiative (WPI) through MEXT, Japan.

%%%%%%%%%%%%%%%%%%%%%%%%%%%%%%%%%%%%%%%%%%%%%%%%%%%%%%%%%%%%%%%%%%%%%%%%%%%%%


\begin{thebibliography}{00}

%% For numbered reference style

\bibitem{Brown:2001nh}
G.~E.~Brown and M.~Rho,
%``On the manifestation of chiral symmetry in nuclei and dense nuclear matter,''
Phys. Rept. \textbf{363}, 85-171 (2002).

\bibitem{Holt:2014hma}
J.~W.~Holt, M.~Rho and W.~Weise,
%``Chiral symmetry and effective field theories for hadronic, nuclear and stellar matter,''
Phys. Rept. \textbf{621}, 2-75 (2016).

\bibitem{Rho:2024ihu}
M.~Rho,
%``The smile of Cheshire Cat at high density,''
J. Subatomic Part. Cosmol. \textbf{1-2}, 100001 (2024).

\bibitem{Rho:2025rqf}
M.~Rho,
%``From nuclear matter with quenched gA to compact-star matter with a signal for emergent hidden scale symmetry,''
J. Subatomic Part. Cosmol. \textbf{4}, 100176 (2025).

\bibitem{Rho:2025mns}
M.~Rho,
%``A Bottom-Up EFT Approach To Superdense Baryonic Matter,''
[arXiv:2511.21141 [nucl-th]].

\bibitem{Fukushima:2013rx}
K.~Fukushima and C.~Sasaki,
%``The phase diagram of nuclear and quark matter at high baryon density,''
Prog. Part. Nucl. Phys. \textbf{72}, 99-154 (2013).



\bibitem{Harada:2000kb}
M.~Harada and K.~Yamawaki,
%``Vector manifestation of the chiral symmetry,''
Phys. Rev. Lett. \textbf{86}, 757-760 (2001).

\bibitem{Harada:2003jx}
M.~Harada and K.~Yamawaki,
%``Hidden local symmetry at loop: A New perspective of composite gauge boson and chiral phase transition,''
Phys. Rept. \textbf{381}, 1-233 (2003).

\bibitem{Harada:2001it}
M.~Harada and C.~Sasaki,
%``Vector manifestation in hot matter,''
Phys. Lett. B \textbf{537}, 280-286 (2002).

\bibitem{Harada:2001qz}
M.~Harada, Y.~Kim and M.~Rho,
%``Vector manifestation and fate of vector mesons in dense matter,''
Phys. Rev. D \textbf{66}, 016003 (2002).



\bibitem{Harada:2003kt}
M.~Harada, M.~Rho and C.~Sasaki,
%``Chiral doubling of heavy light hadrons and the vector manifestation of hidden local symmetry,''
Phys. Rev. D \textbf{70}, 074002 (2004).

\bibitem{Sasaki:2011ff}
C.~Sasaki, H.~K.~Lee, W.~G.~Paeng and M.~Rho,
%``Conformal anomaly and the vector coupling in dense matter,''
Phys. Rev. D \textbf{84}, 034011 (2011).

\bibitem{Paeng:2011hy}
W.~G.~Paeng, H.~K.~Lee, M.~Rho and C.~Sasaki,
%``Dilaton-Limit Fixed Point in Hidden Local Symmetric Parity Doublet Model,''
Phys. Rev. D \textbf{85}, 054022 (2012).

\bibitem{Paeng:2013xya}
W.~G.~Paeng, H.~K.~Lee, M.~Rho and C.~Sasaki,
%``Interplay between $\omega$-nucleon interaction and nucleon mass in dense baryonic matter,''
Phys. Rev. D \textbf{88}, 105019 (2013).



\bibitem{Marczenko:2021uaj}
M.~Marczenko, K.~Redlich and C.~Sasaki,
%``Reconciling Multi-messenger Constraints with Chiral Symmetry Restoration,''
Astrophys. J. Lett. \textbf{925}, no.2, L23 (2022).

\bibitem{Marczenko:2022hyt}
M.~Marczenko, K.~Redlich and C.~Sasaki,
%``Chiral symmetry restoration and {\ensuremath{\Delta}} matter formation in neutron stars,''
Phys. Rev. D \textbf{105}, no.10, 103009 (2022).

\bibitem{Marczenko:2018jui}
M.~Marczenko, D.~Blaschke, K.~Redlich and C.~Sasaki,
%``Chiral symmetry restoration by parity doubling and the structure of neutron stars,''
Phys. Rev. D \textbf{98}, no.10, 103021 (2018).

\bibitem{Marczenko:2020jma}
M.~Marczenko, D.~Blaschke, K.~Redlich and C.~Sasaki,
%``Toward a unified equation of state for multi-messenger astronomy,''
Astron. Astrophys. \textbf{643}, A82 (2020).


\bibitem{Marczenko:2023ohi}
M.~Marczenko, K.~Redlich and C.~Sasaki,
%``Fluctuations near the liquid-gas and chiral phase transitions in hadronic matter,''
Phys. Rev. D \textbf{107}, no.5, 054046 (2023).

\bibitem{Koch:2023oez}
V.~Koch, M.~Marczenko, K.~Redlich and C.~Sasaki,
%``Fluctuations and correlations of baryonic chiral partners,''
Phys. Rev. D \textbf{109}, no.1, 014033 (2024).

\bibitem{Marczenko:2024nge}
M.~Marczenko, K.~Redlich and C.~Sasaki,
%``Probing the nuclear liquid-gas phase transition with isospin correlations,''
Phys. Rev. C \textbf{111}, no.6, 6 (2025).



\bibitem{Nowak:1992um}
M.~A.~Nowak, M.~Rho and I.~Zahed,
%``Chiral effective action with heavy quark symmetry,''
Phys. Rev. D \textbf{48}, 4370-4374 (1993).

\bibitem{Bardeen:1993ae}
W.~A.~Bardeen and C.~T.~Hill,
%``Chiral dynamics and heavy quark symmetry in a solvable toy field theoretic model,''
Phys. Rev. D \textbf{49}, 409-425 (1994).

\bibitem{Nowak:2003ra}
M.~A.~Nowak, M.~Rho and I.~Zahed,
%``Chiral doubling of heavy light hadrons: BABAR 2317-MeV/c**2 and CLEO 2463-MeV/c**2 discoveries,''
Acta Phys. Polon. B \textbf{35}, 2377-2392 (2004).

\bibitem{BaBar:2003oey}
B.~Aubert \textit{et al.} [BaBar],
%``Observation of a narrow meson decaying to $D_s^+ \pi^0$ at a mass of 2.32-GeV/c$^2$,''
Phys. Rev. Lett. \textbf{90}, 242001 (2003).

\bibitem{CLEO:2003ggt}
D.~Besson \textit{et al.} [CLEO],
%``Observation of a narrow resonance of mass 2.46-GeV/c**2 decaying to D*+(s) pi0 and confirmation of the D*(sJ)(2317) state,''
Phys. Rev. D \textbf{68}, 032002 (2003)
[erratum: Phys. Rev. D \textbf{75}, 119908 (2007)].

\bibitem{Belle:2003guh}
P.~Krokovny \textit{et al.} [Belle],
%``Observation of the D(sJ)(2317) and D(sJ)(2457) in B decays,''
Phys. Rev. Lett. \textbf{91}, 262002 (2003).



\bibitem{Detar:1988kn}
C.~E.~Detar and T.~Kunihiro,
%``Linear $\sigma$ Model With Parity Doubling,''
Phys. Rev. D \textbf{39}, 2805 (1989).



\bibitem{LIGOScientific:2018cki}
B.~P.~Abbott \textit{et al.} [LIGO Scientific and Virgo],
%``GW170817: Measurements of neutron star radii and equation of state,''
Phys. Rev. Lett. \textbf{121}, no.16, 161101 (2018).

\bibitem{LIGOScientific:2018hze}
B.~P.~Abbott \textit{et al.} [LIGO Scientific and Virgo],
%``Properties of the binary neutron star merger GW170817,''
Phys. Rev. X \textbf{9}, no.1, 011001 (2019).

\bibitem{Miller:2019cac}
M.~C.~Miller, F.~K.~Lamb, A.~J.~Dittmann, S.~Bogdanov, Z.~Arzoumanian, K.~C.~Gendreau, S.~Guillot, A.~K.~Harding, W.~C.~G.~Ho and J.~M.~Lattimer, \textit{et al.}
%``PSR J0030+0451 Mass and Radius from $NICER$ Data and Implications for the Properties of Neutron Star Matter,''
Astrophys. J. Lett. \textbf{887}, no.1, L24 (2019).

\bibitem{Miller:2021qha}
M.~C.~Miller, F.~K.~Lamb, A.~J.~Dittmann, S.~Bogdanov, Z.~Arzoumanian, K.~C.~Gendreau, S.~Guillot, W.~C.~G.~Ho, J.~M.~Lattimer and M.~Loewenstein, \textit{et al.}
%``The Radius of PSR J0740+6620 from NICER and XMM-Newton Data,''
Astrophys. J. Lett. \textbf{918}, no.2, L28 (2021).

\bibitem{McLerran:2018hbz}
L.~McLerran and S.~Reddy,
%``Quarkyonic Matter and Neutron Stars,''
Phys. Rev. Lett. \textbf{122}, no.12, 122701 (2019).


\end{thebibliography}
\end{document}